\documentclass[11pt,twoside]{article}

\usepackage{asp2006}
\usepackage{epsf}
\usepackage{lscape}

\begin{document}
\title{Progress report: probabilistic and statistical bases of surface brightness fluctuations}   
\author{M. Cervi\~no and V. Luridiana}
\affil{Instituto de Astrof\'\i sica de Andaluc\'\i a (CSIC)}    
\author{and L. Jamet}
\affil{Instituto de Astronom\'\i a (UNAM)}

\begin{abstract} 
The surface brightness fluctuations (SBF) method is a statistical method applied on image pixels in different bands.  
This contribution aims to distinguish between the observational (statistical) method  and the theoretical (probabilistic) method based on stellar population synthesis and needed for the calibration of observational SBF. 
We find that the commonly used SBF theoretical definition as the {\it mean luminosity-weighted luminosity of the stellar population} is only compatible with the observational method under quite strong hypotheses, and that it is not compatible with stellar population theory results.
\end{abstract}


The observational method used to obtain SBF can be described as follows \citep{TS88}. First, an image $f_{x,y}$ of a galaxy is taken, and the {\it local} mean value $\mu_1'(f_{x,y})$ is obtained by an over-smoothing of the original image using a known PSF. Second, the image of fluctuations around the local mean is obtained and divided by the square root of the mean image; this produces an image $f_{x,y}^{\mathrm{fluc}}$. The goal of this step is trying to eliminate the effects of the structure of the galaxy (i.e. the galaxy profile) from the image of fluctuations. Finally, the variance $\mu_2(f^{\mathrm{fluc}})$ of the distribution $\varphi(f^{\mathrm{fluc}})$ is obtained, which describes the values of $f_{x,y}^{\mathrm{fluc}}$ across the image. The observed SBF are defined as the estimate $\bar{f} = \mu_2(f^{\mathrm{fluc}})$. 

In a probabilistic framework, neglecting sampling effects, it is trivial to see that the mean of  $\varphi(f^{\mathrm{fluc}})$, $\mu_1'(f^{\mathrm{fluc}})$, is zero, since the mean of $f_{x,y} - \mu_1'(f_{x,y})$ is zero over any local set of pixels. However, the variance of this distribution is not trivially obtained and depends on additional assumptions.

The variance of $\varphi(f^{\mathrm{fluc}})$ can be obtained if all the {\it possible} values of the integrated luminosity in a pixel are known. This amounts to knowing the population luminosity distribution function (pLDF) for the case of $N^*$ stars: $\varphi_{\mathrm{L_{\mathrm{tot}}}}(L_{\mathrm{theo};N^*})$. \cite{CL06} show that the mean and the variance (($\mu_{1}'[\varphi_{\mathrm{L_{\mathrm{tot}}}}(L_{\mathrm{theo};N^*})]$ and $\mu_{2}[\varphi_{\mathrm{L_{\mathrm{tot}}}}(L_{\mathrm{theo};N^*})]$) of the pLDF can be related through simple relations with the mean and the variance ($\mu_{1}'$ and $\mu_{2}$) of the stellar luminosity distribution function (sLDF, $\varphi_{\mathrm{L}}$, i.e. the pLDF for $N^* =1$) (see also Luridiana \& Cervi\~no in these proceedings):

\begin{eqnarray}
\mu_{1}'[\varphi_{\mathrm{L_{\mathrm{tot}}}}(L_{\mathrm{theo};N^*})] = N^* \times \mu_{1}'   &;&
\mu_{2}[\varphi_{\mathrm{L_{\mathrm{tot}}}}(L_{\mathrm{theo};N^*})] = N^* \times \mu_{2}. \nonumber
\end{eqnarray}

After performing the same set of operations defined by the observational method on the pLDF, the following is obtained:

\begin{equation}
\bar{L}_{\mathrm{theo}} = \frac{\mu_{2}'[\varphi_{\mathrm{L_{\mathrm{tot}}}}(L_{\mathrm{theo};N^*})]}{\mu_{2}'[\varphi_{\mathrm{L_{\mathrm{tot}}}}(L_{\mathrm{theo};N^*})]} = \frac{N^*\, \mu_{2}}{N^* \, \mu_{1}'} =  \frac{\mu_{2}}{\mu_{1}'}.
\label{Eq:SBFteo}
\end{equation}

\noindent This expression does not depend on $N^*$ and should hold in all the galaxy pixels.
This is the theoretical value that can be used to calibrate the SBF. However, in the observational domain, we have no access to the number of stars in individual pixels and it is necessary to estimate a mean value on a given set of pixels. Therefore an additional distribution is needed, $\psi_{\mathrm{PSF}}(N^*)$, which describes the variation of the number of stars in the pixels from which the mean is obtained.

The distribution of possible values of the luminosity in this set of pixels, $\phi_{\mathrm{PSF}}(L_{\mathrm{PSF}})$, is the result of the sum of all possibles $\varphi_\mathrm{L_{tot}}(L_{\mathrm{theo};N^*})$ distributions weighted by the probability that the set of pixels would have a number of stars $N^*$: 
$\phi_{\mathrm{PSF}}(L_{\mathrm{PSF};N_{\mathrm{PSF}}}) = \sum_{N^*} \,\psi_{\mathrm{PSF}}(N^*) \, \varphi_\mathrm{L_{tot}}(L_{\mathrm{theo};N^*})$. 
The mean and variance of this distribution are:

\begin{eqnarray}
\mu_{1}'(L_{\mathrm{PSF}}) &=& \mu_{1}'(N_{\mathrm{PSF}})\, \mu_{1}' \label{eqs:k1PSF}, \\
\mu_{2}(L_{\mathrm{PSF}}) &=& \mu_{1}'(N_{\mathrm{PSF}}) \, \mu_{2}  + \mu_{2}(N_{\mathrm{PSF}})\, \mu_{1}'^2.  \label{eqs:k2PSF}  
\label{eqs:k4PSF}
\end{eqnarray}

And now, the new distribution from the fluctuating image, $\phi_{\mathrm{PSF}}(L_{\mathrm{PSF}}^\mathrm{fluc})$, has a mean value equal to zero and variance (i.e. the SBF) given by:

\begin{equation}
\bar{L}_{\mathrm{PSF}} = \frac{\mu_{2}(L_{\mathrm{PSF}})}{\mu_{1}'(L_{\mathrm{PSF}})} = \bar{L}_{\mathrm{theo}} + \mu_{1}' \frac{\mu_{2}(N_{\mathrm{PSF}})}{\mu_{1}'(N_{\mathrm{PSF}})}.
\label{eq:SBFgeneral}
\end{equation}

If $\psi_{\mathrm{PSF}}(N^*)$ is a Poisson distribution,

\begin{equation}
\bar{L}_{\mathrm{PSF};N_{\mathrm{poi}}} = \bar{L}_{\mathrm{theo}} + \mu_{1}' = \frac{\mu_{2}'}{\mu_{1}'}, 
\label{eq:SBFPoi}
\end{equation}

\noindent where $\mu_{2}'$ is the second raw moment of the sLDF (which is related with the variance and the mean value of the sLDF through $\mu_{2}=\mu_{2}' - \mu_{1}'^2$). Eq. \ref{eq:SBFPoi} is the definition proposed in \cite{TS88} for the theoretical SBF: the luminosity-weighted luminosity of the stellar population.
However, Eq. \ref{eq:SBFPoi} is only valid under the observational hypothesis of a Poisson distribution and if the galaxy has a flat luminosity profile: that is, the definition depends on observational constraints and cannot be used to define theoretical SBF.
We propose to use Eq. \ref{Eq:SBFteo} instead, which can be compared with observations after they are corrected for the PSF effects given in Eq. \ref{eq:SBFgeneral}.

\acknowledgements 
This work was supported by the Spanish {\it Programa Nacional de Astronom\'\i a y Astrof\'\i sica} through the project AYA2004-02703. MC has a {\it Ram\'on y Cajal} fellowship. VL has a {\it CSIC-I3P} fellowship.

\end{document}